\documentstyle[sprocl]{article}

\def\Journal#1#2#3#4{{#1} {\bf #2}, #3 (#4)}

\def\PLB{{\em Phys. Lett.}  B}
\def\PRL{\em Phys. Rev. Lett.}
\def\PRD{{\em Phys. Rev.} D}
\def\ZPC{{\em Z. Phys.} C}

\def\ra{\rightarrow}

\def\vp{{\bf p}}

\def\be{\begin{equation}}
\def\ee{\end{equation}}
\def\bea{\begin{eqnarray}}
\def\eea{\end{eqnarray}}

\begin{document}
\def\be{\begin{eqnarray}}
\def\en{\end{eqnarray}}
\def\non{\nonumber}
\def\la{\langle}
\def\ra{\rangle}
\def\ep{\varepsilon}
\newcommand{\eps}{\varepsilon}
\newcommand{\B}{{\mathcal{B}}}
\newcommand{\F}{{\mathcal{F}}}
\newcommand{\G}{{\mathcal{G}}}
\def\up{\uparrow}
\def\dw{\downarrow}
\def\non{\nonumber}
\def\la{\langle}
\def\ra{\rangle}
\def\nc{N_c^{\rm eff}}
\def\vp{\varepsilon}
\def\vcbud{{V_{cb}V_{ud}^*}}
\def\vcbcs{{V_{cb}V_{cs}^*}}
\def\C1{{\left( \frac{G_F}{\sqrt{2}}\vcbud\right)^2}}
\def\r{{\tau(B^-)\over\tau(\bar{B}^0)}}
\def\lsim{ {\ \lower-1.2pt\vbox{\hbox{\rlap{$<$}\lower5pt\vbox{\hbox{$\sim$}
}}}\ } }
\def\gsim{ {\ \lower-1.2pt\vbox{\hbox{\rlap{$>$}\lower5pt\vbox{\hbox{$\sim$}
}}}\ } }

\font\el=cmbx10 scaled \magstep2

\begin{flushright}
 hep-ph/9809528
\end{flushright}
\vspace{1.0cm}

\title{\bf Updated Analysis of $a_2/a_1$ in Exclusive B Decays
\footnote{presented by K.-C. Yang at ``The Fourth International
Workshop on Particle Physics Phenomenology", Kaohsiung, Taiwan,
18-21 January 1998.\\ Email address: {\tt
kcyang@phys.sinica.edu.tw}}}
\author{Hai-Yang Cheng and Kwei-Chou Yang}
\address{Institute of Physics, Academia Sinica, Taipei, Taiwan 115, R.O.C.}

\maketitle\abstracts{
 Using recent experimental data and various theoretical calculations on form factors,
 we reanalyze the effective parameters  $a_1$ and $a_2$ and their ratio.}
Nonleptonic two-body decays of mesons were conventionally studied
in the generalized factorization approach, in which the factorized
decay amplitude is associated with the effective coefficients
\begin{eqnarray}
a_{1,2}^{\rm eff}= c_{1,2}^{\rm eff} + c_{2,1}^{\rm eff}
[({1/N_c}) +\chi_{1,2}]\,,
\end{eqnarray}
where $c_{1,2}^{\rm eff}$ are the effective Wilson coefficients
and the nonfactorized terms are expressed by $\chi_{1,2}$. For
simplicity, in what follows we will drop the superscript ``eff" of
$a_{1,2}^{\rm eff}$ . Based on the generalized factorization
assumption, one can catalog the decay processes into three
different classes. For class-I decays, the decay amplitudes,
dominated by color-allowed external W-emission, are proportional
to the $a_1$ parameter. For class-II decays, the decay amplitudes,
governed by color-suppressed internal W-emission, are described by
$a_2$. The decay amplitudes of the class-III involve a linear
combination of $a_1$ and $a_2$. From the data analysis, contrary
to the case of charmed meson decays in which $a_2$ is negative and
$\chi \sim -1/N_c $, $a_2$ is positive in  two-body decays of the
$B$ meson. $\chi_{2(1)}$ coming from the contribution of the
matrix element of the color octet-octet and singlet-singlet
currents, since it is made of nonfactorizable soft gluon effects
in itself, will become more important if the final particles move
slower~\cite{ct,cheng}. Therefore, we na\"\i vely expect that
$|\chi_{i}(D\to V~P)|\gsim |\chi_{i}(D\to P~P)|> |\chi_{i}(B\to
V~P)|$, where $P$ is the pseudoscalar meson and $V$ the vector
meson. The data of $B\to J/\Psi~K^{(*)}$ which are class-II decay
modes can be used to study the absolute value of $a_2$. However,
the experimental results are quite difficult to account for the
observed longitudinal polarization fraction $\Gamma_L/\Gamma$ and
the ratio $R\equiv \Gamma(B\to J/\Psi~K^*)/\Gamma(B\to J/\Psi~K)$
simultaneously using the existing form factor calculations. To
accommodate the data, it has been advocated~\cite{cheng} a
modification to the generalized factorization hypothesis so that
$a_2$ is process dependent. In this talk, we focus on the
extraction of the value $a_1$, $|a_2|$, and $a_2/a_1$ from the
current experimental data~\cite{PDG} and various form factor
calculations~\cite{ask}, the BSWI model, the BSWII
model~\cite{bsw}, the LF model~\cite{cch}, the NS model~\cite{ns},
the QCD sum rule results (Yang)~\cite{yang}, and the light-cone
sum rule calculations (LC)~\cite{bb1}. Since the accuracy of these
values still suffers from reliability of the form factor
calculations, we desire that an objective estimation can be
obtained.

Here will consider the decay amplitudes of processes which have
been measured experimentally:
\begin{itemize}
\item{Class I:} $B^- \to D^{(*)0}~D_s^{(*)-}$, $\bar
B_d^0 \to D^{(*)+}~D_s^{(*)-}$, and
 $\bar B_d^0 \to D^{(*)+} ~\pi^-(\rho^-)$\,,
\end{itemize}
\begin{itemize}
\item{Class II:} $B^- \to J/\Psi~ K^{(*)-}$ and  $B^0\to J/\Psi K^{(*)0}$  \,.
\end{itemize}
\begin{itemize}
\item{Class III:} $B^- \to D^{(*)0} ~\pi^-(\rho^-)$\,,
\end{itemize}
The value of $a_{(1,2)}$ can be extracted from the above data of
the class-(I,II) type transitions, while their ratio can be
obtained by the the following ratios: \be\label{R14}
R_1&&\equiv{{\cal B}(B^-\to D^0\pi^-)\over{\cal B}(\bar{B}^0\to
D^+\pi^-)}, \ \ \ \ R_2\equiv{{\cal B}(B^-\to D^0\rho^-)\over{\cal
B}(\bar{B}^0\to D^+\rho^-)}\,,\nonumber\\ R_3&&\equiv{{\cal
B}(B^-\to D^{*0}\pi^-)\over{\cal B}(\bar{B}^0\to D^{*+}\pi^-)}\,,
\ \ R_4\equiv{{\cal B}(B^-\to D^{*0}\rho^-)\over{\cal
B}(\bar{B}^0\to D^{*+} \rho^-)} \,.
\en

\begin{table}[t]\caption{The effective parameter $a_1$ extracted
from $B \to D^{(*)}~D_s^{(*)}$ using different form-factor
models.\label{tab:a1DD}}
\begin{center}
\footnotesize
\begin{tabular}{|cccccc|}
\hline
  &     BSWI     &    BSWII     &      LF       &   SR+HQET    & NS \\
\hline $\bar B^0\to D^+ D_s^-$  & 1.05$\pm$0.16 & 1.05$\pm$0.16 &1.06$\pm$0.16 &
 1.17$\pm$0.18  & 1.32$\pm$0.20
\\ $B^- \to D^0 D_s^-$  & 0.80$\pm$0.15 & 0.80$\pm$0.15 & 0.81$\pm$0.15 &
0.89$\pm$0.17 & 1.01$\pm$0.19 \\ \hline $\bar B^0\to D^+
D_s^{*-}$&1.24$\pm$0.31&1.10$\pm$0.27&1.11$\pm$0.28&1.26$\pm$0.31&1.43$\pm$0.36\\
$B^- \to D^0 D_s^{*-}$ &1.14$\pm$0.25 & 1.10$\pm$0.22
&1.02$\pm$0.23&1.16$\pm$0.26&1.32$\pm$0.29\\ \hline $\bar B^0\to
D^{*+} D_s^-$  &1.34$\pm$0.24 & 1.21$\pm$0.21 & 1.11$\pm$0.28 &
1.09$\pm$0.19&1.03$\pm$0.18\\ $B^- \to D^{*0}D_s^-$  &
1.46$\pm$0.30 & 1.31$\pm$0.28 & 1.02$\pm$0.23 & 1.18$\pm$0.25  &
1.12$\pm$0.23 \\ \hline $\bar B^0\to D^{*+} D_s^{*-}$  &
0.88$\pm$0.15 & 0.90$\pm$0.16 &0.83$\pm$0.15 & 0.86$\pm$0.15  &
0.85$\pm$0.15 \\ $B^- \to D^{*0} D_s^{*-}$  & 1.05$\pm$0.20 &
1.07$\pm$0.20 & 0.99$\pm$0.18 &1.03$\pm$0.19  & 1.02$\pm$0.19 \\
\hline Average & 1.03$\pm$0.07 &1.01$\pm$0.07 & 0.97$\pm$0.07 &
1.03$\pm$0.07  & 1.06$\pm$0.07 \\ \hline
\end{tabular}
\end{center}
\end{table}
\begin{table}[ht] \caption{The effective parameter $a_1$ extracted
from $\bar B_d^0 \to D^{(*)+} ~\pi^-(\rho^-)$ using different
form-factor models. \label{tab:a1Dpi}}
 \begin{center}
\footnotesize
\begin{tabular}{|cccccc|}
\hline
                             &     BSWI     &    BWSII     &      LF       &
                              SR+HQET   & NS \\
\hline
   $\bar B^0\to D^+ \pi^-$   & 0.89$\pm$0.06 & 0.89$\pm$0.06 & 0.88$\pm$0.06 &
    1.15$\pm$0.08 & 1.31$\pm$0.09 \\
  $\bar B^0\to D^+ \rho^-$   & 0.91$\pm$0.08 & 0.90$\pm$0.08 & 0.89$\pm$0.08 &
   1.13$\pm$0.10 & 1.29$\pm$0.11 \\
 $\bar B^0\to D^{*+} \pi^-$  & 0.98$\pm$0.04 & 0.98$\pm$0.04 & 0.84$\pm$0.03 &
  0.93$\pm$0.04 & 0.89$\pm$0.03 \\
 $\bar B^0 \to D^{*+}\rho^-$ & 0.86$\pm$0.21 & 0.87$\pm$0.21 & 0.75$\pm$0.18 &
  0.83$\pm$0.20 & 0.80$\pm$0.20 \\
\hline
          Average            & 0.95$\pm$0.03 & 0.94$\pm$0.03 & 0.85$\pm$0.03 &
           0.98$\pm$0.03 & 0.96$\pm$0.03 \\
\hline
\end{tabular}
\end{center}
\end{table}
\begin{table}[t] \caption{The effective parameter $|a_2|$ extracted
from $B\to J/\Psi~ K^{(*)}$ using different form-factor models.
\label{tab:a2}}
\begin{center}
\footnotesize
\begin{tabular}{|ccccccc|}
\hline
                        &     BSWI     &    BSWII     &      LF       &
                              NS       &   Yang   & LC  \\
\hline
  $B^+\to J/\Psi K^+$   & 0.33$\pm$0.02 & 0.22$\pm$0.01 & 0.28$\pm$0.01 &
  0.36$\pm$0.02 & 0.37$\pm$0.02 & 0.30$\pm$0.02 \\
  $B^0\to J/\Psi K^0$   & 0.32$\pm$0.02 & 0.21$\pm$0.01 & 0.27$\pm$0.02 &
   0.35$\pm$0.02 & 0.36$\pm$0.02 & 0.29$\pm$0.02 \\
        Average         & 0.32$\pm$0.01 & 0.22$\pm$0.01 & 0.28$\pm$0.01 &
         0.36$\pm$0.01 & 0.36$\pm$0.01 & 0.30$\pm$0.01 \\
\hline
 $B^+\to J/\Psi K^{*+}$ & 0.20$\pm$0.02 & 0.21$\pm$0.02 & 0.25$\pm$0.02 &
 0.24$\pm$0.02 & 0.39$\pm$0.04 & 0.20$\pm$0.02 \\
 $B^0\to J/\Psi K^{*0}$ & 0.19$\pm$0.01 & 0.21$\pm$0.01 & 0.25$\pm$0.02 &
  0.25$\pm$0.02 & 0.39$\pm$0.03 & 0.19$\pm$0.01 \\
        Average         & 0.20$\pm$0.01 & 0.22$\pm$0.01 & 0.26$\pm$0.01 &
         0.25$\pm$0.01 & 0.40$\pm$0.02 & 0.20$\pm$0.01 \\
\hline
\end{tabular}
\end{center}
\end{table}
\begin{table}[ht]
\caption{R, $\Gamma_L/\Gamma$, and $|P|^2$  from various form
factor calculations based on the generalized factorization
hypothesis together with the CLEO new data.\label{tab:rgp}}
\begin{center}
\footnotesize
\begin{tabular}{|cccccccc|}
\hline
                 & BSWI &BSWII &  LF  & NS   & Yang & LC   &CLEO\\
\hline
  R              & 4.15  & 1.58  & 1.79 & 3.17 & 1.30      & 3.48 & $1.45\pm 0.26$\\
$\Gamma_L/\Gamma$& 0.57  & 0.36  & 0.53 & 0.49 & 0.42      & 0.47
& $0.52\pm 0.08$\\
 $|P|^2$         & 0.09  & 0.24  & 0.09 & 0.12 & 0.19      & 0.23 & $0.16\pm 0.09$\\
\hline
\end{tabular}
\end{center}
\end{table}
\begin{table}[ht]
\caption{$a_2/a_1$ extracted from the PDG data.
\label{tab:a2a1pdg}}
\begin{center}
\footnotesize
\begin{tabular}{|ccccccc|}
\hline
       &     BSWI     &    BSWII     &      LF       &      NS       & Yang &  LC   \\
\hline
 $R_1$ & 0.29$\pm$0.10 & 0.29$\pm$0.10 & 0.38$\pm$0.13 & 0.28$\pm$0.10 &
   0.29$\pm$0.10   & 0.27$\pm$0.10  \\
 $R_2$ & 0.59$\pm$0.30 & 0.52$\pm$0.27 & 0.56$\pm$0.28 & 0.43$\pm$0.22 &
  1.21$\pm$0.62   & 0.34$\pm$0.17 \\
 $R_3$ & 0.23$\pm$0.06 & 0.20$\pm$0.06 & 0.31$\pm$0.09 & 0.30$\pm$0.09 &
   0.28$\pm$0.08   & 0.25$\pm$0.07  \\
 $R_4$ & 0.55$\pm$0.40 & 0.67$\pm$0.49 & 0.85$\pm$0.62 & 0.79$\pm$0.58 &
  1.50$\pm$1.09   & 0.62$\pm$0.45  \\
 Average & 0.26$\pm$0.05& 0.23$\pm$0.05 & 0.35$\pm$0.07 & 0.31$\pm$0.06 &
   0.30$\pm$0.06    & 0.27$\pm$0.05  \\
\hline
\end{tabular}
\end{center}
\end{table}
\begin{table}[ht]
\caption{$a_2/a_1$, if using the CLEO II data.
\label{tab:a2a1cleo}}
\begin{center}
\footnotesize
\begin{tabular}{|ccccccc|}
\hline
       &     BSWI     &    BSWII     &      LF       &      NS       & Yang &  LC   \\
\hline
 $R_1$ & 0.33$\pm$0.13 & 0.33$\pm$0.13 & 0.43$\pm$0.16 & 0.32$\pm$0.12 &
  0.33$\pm$0.13   & 0.31$\pm$0.12  \\
 $R_2$ & 0.10$\pm$0.24 & 0.09$\pm$0.21 & 0.09$\pm$0.23 & 0.07$\pm$0.18 &
  0.20$\pm$0.49   & 0.06$\pm$0.14 \\
 $R_3$ & 0.16$\pm$0.08 & 0.14$\pm$0.07 & 0.22$\pm$0.10 & 0.22$\pm$0.10 &
  0.20$\pm$0.09   & 0.18$\pm$0.08  \\
 $R_4$ & 0.33$\pm$0.14 & 0.40$\pm$0.18 & 0.51$\pm$0.22 & 0.47$\pm$0.21 &
  0.90$\pm$0.39   & 0.37$\pm$0.16  \\
 Average & 0.22$\pm$0.06 & 0.20$\pm$0.05 & 0.29$\pm$0.08 & 0.25$\pm$0.07 &
  0.27$\pm$0.07   & 0.21$\pm$0.06  \\
\hline
\end{tabular}
\end{center}
\end{table}

The hadronic matrix element of effective operators $O_{1,2}$ can
be redefined in the scheme- and scale($\mu$)-dependent ways $
\langle O(\mu)\rangle=g(\mu)\langle O\rangle_{\rm tree}\,, $ where
$\langle O\rangle_{\rm tree}$ is $\mu$-independent. We thus can
rewrite the matrix element of the effective Hamiltonian $\langle
H_{\rm eff}\rangle$ as the product of $c^{\rm eff}$  and $\langle
O\rangle_{\rm tree}$, both of which are scheme- and
$\mu$-independent. The only relevant scale, which is implicit and
of order $m_b$, separating $c^{\rm eff}$ and $\langle
O\rangle_{\rm tree}$ is the energy release of B decays. Here we
use
$
c_1^{\rm eff}=1.149$ and $c_2^{\rm eff}=-0.325\,, $ to the
next-to-leading order~\cite{ct1998}. Using the theoretical results
of form factors and the experimental data from the Particle Data
Group (PDG)~\cite{PDG}, one can easily extract the values of
$a_1$, $|a_2|$, and $a_2/a_1$. The results are listed in Tables
\ref{tab:a1DD}, \ref{tab:a1Dpi}, \ref{tab:a2}, and
\ref{tab:a2a1pdg}. In Tables \ref{tab:a1DD} and \ref{tab:a1Dpi},
the updated data are used for the parameters of NS, while the
results of SR+HQET~\footnote{We will use the  $B\to D^{(*)}$
form-factor results of SR+HQET as subsidiaries of the
Yang's~\cite{yang} and LC~\cite{bb1} calculations, since  Yang's
sum rules and LC only show the form factors of the $B$ meson into
a light meson.} are gotten by the NS formula but the parameters
are obtained from the sum rule results and the predictions of the
heavy quark effective theory~\cite{neubert}. All the
results~\footnote{The class-I transitions determine the absolute
value of $a_1$. However, since this kind of processes is color
allowed, we expect the sign of $a_1$ is the same as that of
$c_1^{\rm eff}$ and is therefore positive.} in Tables
\ref{tab:a1DD} and \ref{tab:a1Dpi} indicate that the value of
$a_1$, which is very close to 1, seems to follow the pattern
$
a_1(B\to D^{(*)} D_s^{(*)})\gsim a_1(B\to D^{(*)} \pi(\rho))\,.
$

For $B\to J/\Psi~ K^{(*)}$, since they are color suppressed and
sensitive to the nonfactorizable contributions, therefore, they
are very good examples to explore the generalized factorization
hypothesis to see if $a_2$ is universal. On the other hand, since
the energy release in $B\to J/\Psi~ K$ is close to that in $B\to
J/\Psi~ K^{*}$, we expect that the extracted values of $|a_2|$
from these two kinds of processes can be slightly different (the
so-called minimal modified factorization hypothesis). However,
from Table \ref{tab:a2}, we know that only BSWII, LF, and Yang's
sum rules can satisfy our expectation.

The other ways to examine the generalized factorization hypothesis
and/or the quality of various form factor results are to study the
production ratio $ \label{R} R=\B{(B\to J/\Psi~ K^*)}/\B(B\to
J/\Psi~ K)\,,
$
the fraction of longitudinal polarization $\Gamma_L/\Gamma$ , and
the $P-$wave transverse polarization $|P|^2$ measured in the
transversity basis in $B\to J/\Psi~ K^*$ decays. The predictions
from various form factor calculations based on the generalized
factorization hypothesis together with the CLEO new
data~\cite{cleo1997} are shown in Table \ref{tab:rgp}. We find
again that only BSWII, LF, and Yang's sum rules  can accommodate
the data, expect that the value of $\Gamma_L/\Gamma$ in BSWII is a
little smaller than the CLEO data. Furthermore, if allowing the
nonfactorizable contribution $\chi_2$ to be slightly different in
$B\to J/\Psi~K^{(*)}$ and in $B\to D^{(*)}~\pi(\rho)$, we find
that the values of $a_2(B\to J/\Psi~K^{(*)})$ of BSWII/LF are
consistent with that obtained from the analyses of $R_{1-4}$ (see
Tables \ref{tab:a2a1pdg} and \ref{tab:a2a1cleo}) if assuming
$a_1\approx 1$. However, note that the Yang's sum rule results do
not behave like the BSWII or LF model. The value of $a_2$ in the
sum rule analysis of Yang is larger. There are two possibilities
to explain it. One is $\chi_2\approx 0.3$, a lager value compared
to that in BSWII or LF ($\chi_2\approx$0.14 in BWSII, 0.18 in LF
if we have adopted $a_2>0$).  The other possibility is both the
nonfactorizable values $\chi_2$ and $a_2$ become negative in $B\to
J/\Psi~K^{(*)}$, as the D meson decays. The QCD sum rule
calculation on the nonfactorizable contribution in $B\to J/\Psi~K$
was reported to be negative~\cite{kr}, contrary to the generalized
nonfactorization hypothesis. However, it is difficult to
understand why $a_2$ is negative in $B\to J/\Psi~K^{(*)}$ while it
becomes positive in $B^-\to D^{(*)}~\pi^-(\rho^-)$. One way to
solve directly this problem is to evaluate the class-III or
class-II decay channels of $B\to D~ \pi(\rho)$, within framework
of the QCD sum rules to see if $\chi_2$ becomes positive.

From the analyses of $R_{1-4}$ defined in Eq.~(\ref{R14}) one can
obtain $a_2/a_1$. Since we have adopted that $a_1$ is positive, we
can thus determine the sign of $a_2$. The results are listed in
Table \ref{tab:a2a1pdg} if using PDG data. All of the results show
that $a_2$ is positive. Two remarks are in order. First, $a_2/a_1$
of $R_{1,3}$ for all of form factor results are quantitatively
stable, while that of $R_{2,4}$ have large central values and
errors. Second, the prediction of $a_2/a_1$ in LF is slightly
large is because $a_1$ in LF is a smaller value (see Tables
\ref{tab:a1DD} and \ref{tab:a1Dpi}).

CLEO II~\cite{ro} has also reported the data on various class-III
and class-I branching ratios of $B\to D^{(*)}~ \pi(\rho)$. We show
that the analysis results of $a_2/a_1$ in Table
\ref{tab:a2a1cleo}. But now the values of $a_2/a_1$ from $R_{2,4}$
is much smaller the corresponding values in Table
\ref{tab:a2a1pdg}. Eventually, if we account of the fact that
$a_1$ is little small in LF, then $a_2$ is lying in $0.2-0.3$,
which is a reasonable result for all of form factor calculations.

To conclude, we have used the current  experimental data and
various theoretical results of form factors to analyze the
effective coefficients $a_1$ and $a_2$. Our results have shown
that if allowing a minimal modification to the generalized
factorization hypothesis, $i.e.$, the nonfactorizable contribution
$\chi_{1,2}$ can be slightly different in different decay
processes, then we have the following conclusions: (1) $a_1\approx
1$ and
$
a_1(B\to D^{(*)} D_s^{(*)})\gsim a_1(B\to D^{(*)}
\pi(\rho))\,.\nonumber
$
(2) Only the BSW II model, the light-front model, and Yang's sum
rules can satisfy this factorization assumption in $B\to
J/\Psi~K^{(*)}$ analyses. The results are shown in Tables
\ref{tab:a2} and \ref{tab:rgp}. (3) From the results of $a_2/a_1$
listed in Tables~\ref{tab:a2a1pdg} and \ref{tab:a2a1cleo}, we
obtain that $a_2$ is lying in $0.2-0.3$, which is a reasonable
result for all of form factor calculations in the $B\to D^{(*)}~
\pi(\rho)$ decay processes.

This work was supported in part by the National
Science Council of R.O.C. under Grant No. NSC88-2112-M-001-006.

\end{document}